\documentclass[preprint,authoryear,12pt]{elsarticle}
\makeatletter
\def\ps@pprintTitle{%
 \let\@oddhead\@empty
 \let\@evenhead\@empty
 \def\@oddfoot{}%
 \let\@evenfoot\@oddfoot}
\makeatother

\usepackage{graphicx}
\usepackage{url}
\usepackage{amssymb}

\journal{New Astronomy}

\begin{document}

\begin{frontmatter}



\title{Apex determination and detection of stellar clumps in the open cluster M~67}


\author{S.V.~Vereshchagin\footnote{E-mail: svvs@ya.ru~(S.~V.~Vereshchagin); chupina@inasan.ru~(N.~V.~Chupina); devesh@aries.res.in~(Devesh P. Sariya); rkant@aries.res.in (R. K. S. Yadav) and brij@aries.res.in (Brijesh Kumar)}, N.V.~Chupina$^1$, Devesh P. Sariya$^2,^3$, R. K. S. Yadav$^2$ and Brijesh Kumar$^2$}

\address{$^{1}$Institute of Astronomy Russian Academy of Sciences (INASAN),48 Pyatnitskaya st., Moscow, Russia\\
$^{2}$Aryabhatta Research Institute of Observational Sciences, Manora Peak Nainital 263 002, India; Tel: 0091 05942 235583; Fax No: 0091 05942 233439\\
$^{3}$School of Studies in Physics \& Astrophysics, Pt. Ravishankar Shukla University, Raipur-492 010 (CG), India\\
}
\begin{abstract}

We determined the cluster apex coordinates, studied the substructures and performed membership analysis 
in the central part $(34'\times33')$ of the open cluster M~67. We used 
the individual stellar apexes method developed earlier and classical technique of proper 
motion diagrams in coordinate system connected with apex. The neighbour-to-neighbour distance 
technique was applied to detect space details. The membership list was corrected and some 
stars were excluded from the most probable members list.  The apex coordinates have been determined 
as: $A_0=132.97^\circ\pm0.81^\circ$ and $D_0=11.85^\circ\pm0.90^\circ$. The 2D-space star 
density field was analysed and high degree of inhomogeneity was found.

\end{abstract}

\begin{keyword}
Astrometry - open clusters: general - open cluster: individual M 67.


\end{keyword}

\end{frontmatter}


\section{Introduction}
\label{Intro}

M~67 is one of the most studied open clusters among the known open clusters with ages comparable 
to or older than the Sun. M~67 is of Solar metallicity, relatively nearby and has low 
interstellar reddening. It has been comprehensively studied by many authors to establish 
astrometric membership \citep{Sanders1977, Girard1989,Yadav2008}.
Many photometric studies \citep{Montgomery1993, Sandquist2004} 
and rather precise radial velocity and binary search study \citep{Mathieu1986, Melo2001, Pasquini2012} 
have been conducted for the cluster. 
The fundamental parameters along with absolute proper motions have been listed in Table 1.
M~67 moves far above the 
galactic disk on latitude $b=+31.91^\circ$, on $z=830\sin(31.91^\circ)=440$~pc along quite a 
circular orbit and interacts with spiral density waves, which can initiate star formation. Using 
2MASS $JHK$ photometry, \citet{Sarajedini2009} suggested an age of 3.5 Gyr using two 
different theoretical isochrones. 


\begin{table}
\caption{The fundamental parameters for M~67}
\vspace{0.5cm}
\centering
\scriptsize
\begin{tabular}{lll}
\hline\hline
Parameters&Value &Ref.\\
\hline
\noalign{\smallskip}
Equatorial coordinate (J2000)&$\alpha=8.855^h$, $\delta=11.8^\circ$ & WEBDA\\
Galactic coordinate          &$l=215.69^\circ$, $b=+31.91^\circ$& WEBDA\\
The core radius              &$r_c=0.12^\circ$&Kharchenko et al. (2005)\\
The cluster radius           &$r_{cl}=0.78^\circ$ &Kharchenko et al. (2005)\\
Distance from the Sun        &830 pc& \citet{Allen1973} \\
Age                          &3.5--4.0 Gyr&\citet{Sarajedini2009} \\
$[$Fe/H$]$                   &+0.03$\pm$0.01&\citet{Randich2006}\\
Absolute proper motion       &$\mu_\alpha\cos\delta=-9.6\pm1.1,\, \mu_\delta=-3.7\pm0.8$ mas/yr &\citet{Bellini2010} \\
\hline
\label{tab1}
\end{tabular}
\end{table}

\citet{Chupina1998} have detected several stellar clumps inside the low density extended corona 
of M~67 with the help of nearest neighbour distance (NND) method, while the cause of origin 
of these substructures remains unclear. 
Availability of new proper motions and radial velocities data 
for M~67 prompted us to revise the membership of stars in the cluster
with our methods using radial velocities and proper motion data to ascertain membership. 

The stellar apexes diagram or $AD$-diagram, \citep{Chupina2001} is useful for the investigation of 
kinematic structures of the star clusters and streams. 
It allows us to find the kinematic substructures inside these objects. 
The AD-diagram is the plot of the individual star apexes. 
The individual apexes represent the equatorial 
coordinates of the point on the celestial sphere 
in which the space velocity vector intersects it.
By analogy, the star apex coordinates of the cluster 
are designated in equatorial coordinates as $A$ for right ascension and $D$ for declination. 
The formal description of this method and the formulae of the error ellipses 
are given in \citet{Chupina2001}. 
It should be noted that the error ellipses can be constructed 
only using the Hipparcos data because it 
contains the necessary correlation coefficients.

The main purpose of the present analysis is the membership revision
of \citet{Yadav2008} catalogue
with the help of convergent point method.
For this purpose, we determined apex coordinates
and used the methods from our previous work \citep{Chupina2001}.
The new data has also been used to study the substructures in its central part.
This work complements our previous one on the corona of the cluster \citep{Chupina1998}.


\begin{table}
\caption{Characteristics of the data taken from \citet{Yadav2008} catalogue.}
\vspace{0.5cm}
\centering
\begin{tabular}{ccccc}
\hline\hline
V range& $\sigma_{\mu_{\alpha}}$& $\sigma_{\mu_{\delta}}$ & $\sigma_{V_r}$ & N\\
\hline
\noalign{\smallskip}
7-10    & 1.86 & 3.27 & 0.008 & 3 \\
10 - 13 & 2.48 & 2.61 & 0.099 & 120 \\
13 - 16 & 2.24 & 2.99 & 0.156 & 451 \\
16 - 19 & 3.60 & 4.13 & 0.116 & 570 \\
19 - 22 &15.16 &15.52 & --    & 1266 \\
\hline
\label{tab2}
\end{tabular}
\end{table}

The structure of the article is as follows:
The data used for the present analysis is described in Sect.~2,
while Sect.~3 is devoted to apex determination. Comments on membership are presented in Sect.~4, 
while Sect.~5 describes the substructures in the central part of the cluster. 
Finally, in Sect. 6 we list the conclusions of the present analysis. 

\section{Data used}

For the present analysis, we used the proper motion catalogue provided by \citet{Yadav2008} 
(hereafter, Yadav08). 
Yadav08 catalogue is one of the richest by number of stars. 
The particulars of this catalogue have been listed in Table~\ref{tab2}.
It contains the median values of errors in proper motion and radial velocity 
with number of stars in different magnitude bins.

The size of the sky area covered and the star 
density allows us to study the cluster nucleus.
It contains relative proper motions and membership probabilities for the 
stars laying towards the central region of the cluster. The catalogue includes the stars
brighter than $V\sim22$ mag in the area of $34^\prime\times33^\prime$.


\begin{table*}
\scriptsize
\caption{The results of the cross-identification
 between  Yadav08 and Hipparcos catalogues}
\vspace{0.5cm}
\label{tab3}
\centering
\tabcolsep=0.5mm
\begin{tabular}{c r l l c c c c c c c}
\hline\hline
\noalign{\smallskip}
Catalog&Star No.&$\alpha$&$\delta$&$B$&$V$&$P$ &$V_r$&$\sigma_{V_r}$ &$\pi$&$\sigma_{\pi}$\\
&&(deg)&(deg)&(mag)&(mag)&(\%)&(km/s)&(km/s)&(mas)&(mas)\\
\hline
\noalign{\smallskip}
 Yadav08&  813& 132.799075  & 11.756134  & 9.872& 10.037 & 98& -28.279& 0.260&     &     \\
 \citet{Hipparcos1997}&43465& 132.79910477& 11.75615394& 9.868& 10.030  &   &        &      & 1.05& 1.96\\
 \citet{vanLeeuwen2005}&43465&             &            &      & 10.012&   &        &      & 1.94& 2.29\\
\hline
\noalign{\smallskip}
 Yadav08& 1024& 132.874609  & 11.788015  & 11.029& 9.586 & 98& 31.655& 0.005&      &     \\
 \citet{Hipparcos1997}&43491& 132.87463223& 11.78802421& 11.269& 9.690  &   &       &      & -1.21& 1.88\\
\citet{vanLeeuwen2005}&43491&             &            &       & 9.818&   &       &      & -0.73& 1.81\\
\hline
\end{tabular}
\end{table*}

Data used to determine the proper motion and membership probability in Yadav08 catalogue were 
taken with Wide-Field-Imager (WFI) mounted on MPG/ESO 2.2~m telescope located at La Silla, Chile. 
There are few blank strips present in the observed region of the 
cluster. These are the gaps between the CCD chips. The catalogue and our method of astrometric 
analysis allowed us to investigate the membership and to obtain some characteristics of the cluster 
space motion and its internal morphology.

The relative proper motions and their errors were determined for $\sim$ 2400 stars using the technique 
described in \citet{Anderson2006} for WFI images. This technique has been used for many star 
clusters to find out relative proper motions of the stars within the cluster region 
\citep{Yadav2008, Sariya2012, Yadav2013}. The accuracy of proper 
motions are from 1.9~mas/yr (for the optimal expositions) to 5.0~mas/yr (for the faintest stars). 

The radial velocity for 211 stars (up to $V\sim16$ mag)
are also provided in the catalogue determined using 
archival VLT-stars spectra. The spectroscopic data were reduced using the GIRAFFE pipeline 
GIRBLDRS \citep{Blecha2000}, in which the spectra were de-biased, flat-field corrected 
and wavelength calibrated, using both prior and simultaneous calibration-lamp spectra. To 
measure the heliocentric radial velocity, they used ``gyCrossC.py'' utility of the GIRAFFE 
pipeline. The formal errors reported in the catalogue are merely the output of pipeline, 
which is clearly underestimating the true errors. Nevertheless, these estimates retain the 
information on the goodness of the fit of the cross-correlation function. 

The error characteristics for relative proper motion and radial velocity are given in 
Table~\ref{tab2}. 

In order to check the parallax values for stars in Yadav08 catalogue, we have carried out the 
cross-identification between Yadav08 and Hipparcos catalogue \citep{Perryman1997}.
We found only two stars, which are listed in Table~\ref{tab3}. 
The large errors in parallaxes prevent us from using them 
for the further analysis. We also note that star number 813 has 
negative $V_r$ value, while its membership probability 
is $P$=98\%. We will discuss about it in the next section.

\begin{figure}[]
\centering
\vspace{-0.7cm}
\includegraphics[width=14cm]{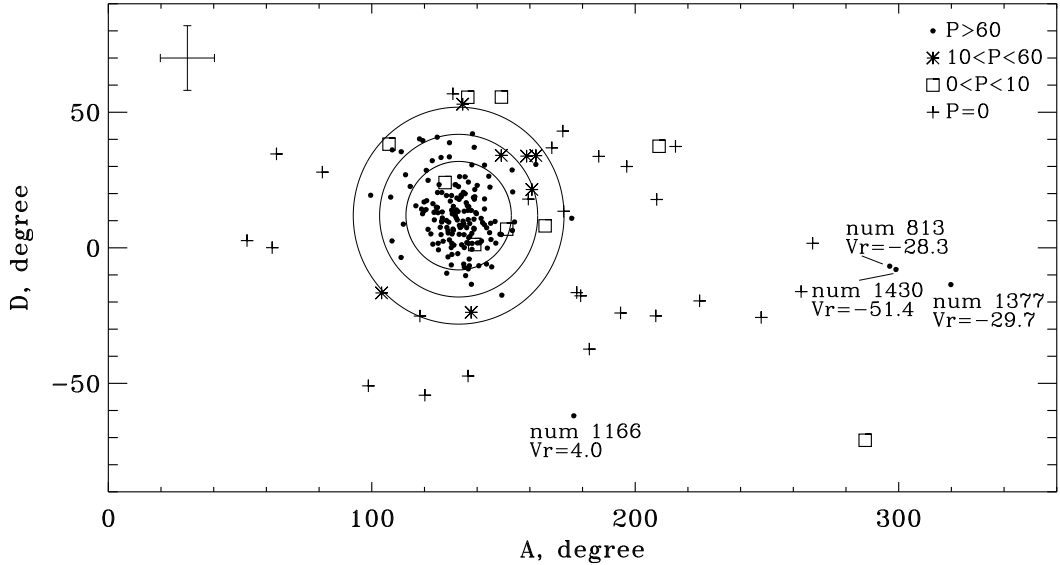}
\caption{The AD-diagram for 211 stars with known radial velocities and different membership 
probability ranges, as shown by different symbols. The circles have the radii 20, 30 and 40 (in degrees) 
respectively. 
Stars for which $V_r$ are considerably different 
from the mean value for the cluster are signed 
with Yadav08 numbers and $V_r$-value.
Details about these stars are in the text.
The average bars of A, D-errors are shown in the top left corner.
}
\label{fig1}
\end{figure}

We have calculated the rms errors of the individual apexes by error propagation formula: 
$\varepsilon^2_A=\sum\limits_q\varepsilon^2_q\times\left(\frac{\partial A}{\partial q}\right)^2_{q=\mu_\alpha,\mu_\delta,V_r}$, 
$\varepsilon^2_D=\sum\limits_q\varepsilon^2_q\times\left(\frac{\partial D}{\partial q}\right)^2_{q=\mu_\alpha,\mu_\delta,V_r}$. 
The average rms errors of A and D 
($\left<\varepsilon_A\right>=10.27^\circ$,  
$\left<\varepsilon_D\right>=11.92^\circ$) 
are shown on Figure~\ref{fig1}. 
The values of average rms errors for stars in AD-area are: 
$\left<\varepsilon_A\right>=9.59^\circ$, 
$\left<\varepsilon_D\right>=11.32 ^\circ$. 
A total of 206 stars have been used for the calculation purpose with rms errors less than $50^\circ$. 

\section{The apex determination}

Yadav08 determined membership probabilities using relative proper motions. 
In this catalogue, we want to refine their membership list. 
To select members of the cluster, we use the proper motions in 
the coordinate system centered at each star and oriented to the convergent point. One of the axes of 
this coordinate system is directed from the star to the cluster apex (U-axis), another is 
perpendicular to it and has the positive direction to the North Pole (T-axis) \citep{vanAltena1969}. 
The concentration of the points on the ``$\mu_U-\mu_T$''-diagram identifies the membership probabilities 
of the cluster. The ``$\mu_U-\mu_T$''-diagram is preferable than ``$\mu_{\alpha}-\mu_{\delta}$'' as it 
excludes the geometric effect of proper motions vector convergence to the apex. 

To construct the ``$\mu_U-\mu_T$''-diagram, we need to know the coordinates of cluster apex. For apex 
determination, we used the AD-diagram method \citep{Chupina2001}, which is very useful for the 
membership list correction too. 
In the AD-diagram, the individual star apexes are plotted, which are 
calculated using coordinates, proper motions, parallax and radial velocity. Unfortunately, we do not 
have individual parallax values for our data. For this reason, we are compelled to use a particular 
distance value (830 pc) for all the stars equal to the distance to the cluster center.


\begin{table}
\caption{The dependence of cluster apex coordinates using different cluster distance values from the 
Sun}
\label{tab4}
\centering
\begin{tabular}{c c c}
\hline\hline
\noalign{\smallskip}
Distance&$A_0$&$D_0$\\
(pc) & (deg)        & (deg) \\
\hline
\noalign{\smallskip}
700&132.95&11.85\\
830&132.97&11.85\\
900&132.98&11.86\\
\hline
\end{tabular}
\end{table}
The choice of distance from the Sun to the cluster center is important. In literature, we found a wide 
range of distances from 800 to 900~pc using different methods. In \citet{Yakut2009}, the distance 
of M~67 was determined as 857~pc via binary star parameters, while \citet{Sarajedini2009} estimated 
the distance as 870~pc with the help of deep near-IR color-magnitude diagram. \citet{Majaess2011} 
used deep infra-red ZAMS fits and precise Hipparcos parallaxes (d$\le$25~pc) are applied to establish 
distances for several open clusters and for M~67, distance is estimated within 815-840~pc. As shown in  
Table~\ref{tab4}, the effect of distance on the determination of cluster apex is not critical. For this 
reason and to compare the present results with our previous ones \citep{Chupina1998}, we accepted 
the distance value as 830~pc \citep{Allen1973}. 

Figure~\ref{fig1} represents the AD-diagram for M~67. We see in the diagram that most of the stars are
located around the cluster apex position. This implies that they have parallel spatial vectors 
(within the errors). Most stars with high membership probabilities lie 
in region with the radius of $24^\circ$. 
The mean membership probability is around 96\% in this region. 
The stars with smaller probabilities jumped 
far from the apex position.
The most dense concentration of points on the AD-diagram designates the average direction of motion 
of the cluster stars (near cluster apex). Stars most likely to be the members are located close to 
the apex, while the less probable members move away from the apex. However, the scatter of points 
around the cluster apex is seen due to the measurement errors in proper motions and radial velocities. 
Therefore, one can 
not say anything about the probability of membership using AD-diagram.

\begin{figure}[]
\centering
\includegraphics[width=10cm,height=10cm]{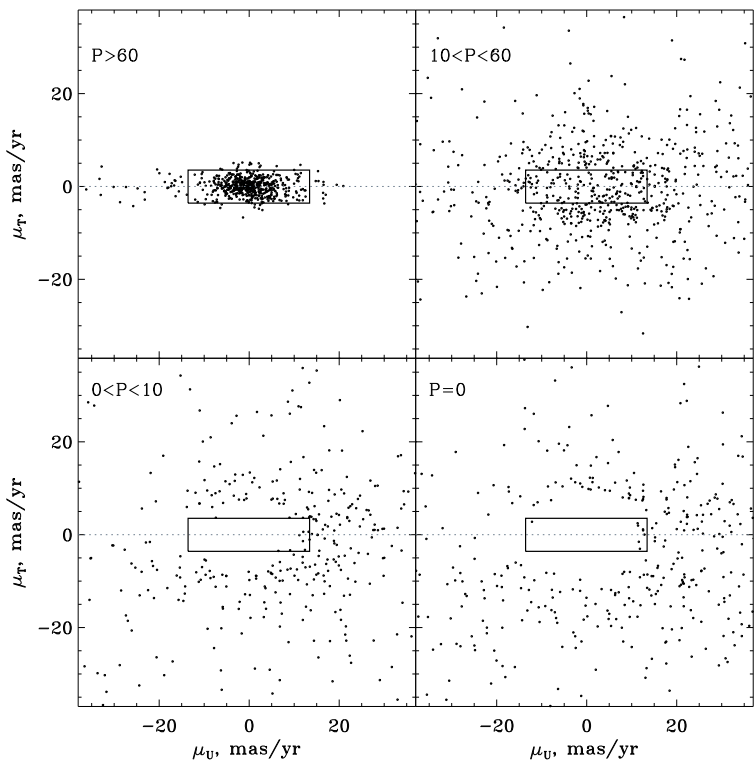}
\caption{The cluster members selection in the ``$\mu_U - \mu_T$''-diagram. 
Panels correspond to the mentioned membership probabilities $P$. 
The rectangular box inside a panel shows $2\sigma$ dispersions 
of $\mu_U$ and $\mu_T$ ($\mu$-box).
}
\label{fig2}
\end{figure}

The AD-diagram allows us to determine the apex position and to correct the membership list. 
We note that the position of four stars with $P>60\%$ 
(with star numbers 1166, 813, 1377, 1430) in Fig.~\ref{fig1} 
is far from the place of maxima of points density. 
Particulars for these stars are listed in Table~\ref{tab6}.

\begin{figure}[]
\centering
\includegraphics[width=10cm,height=10cm]{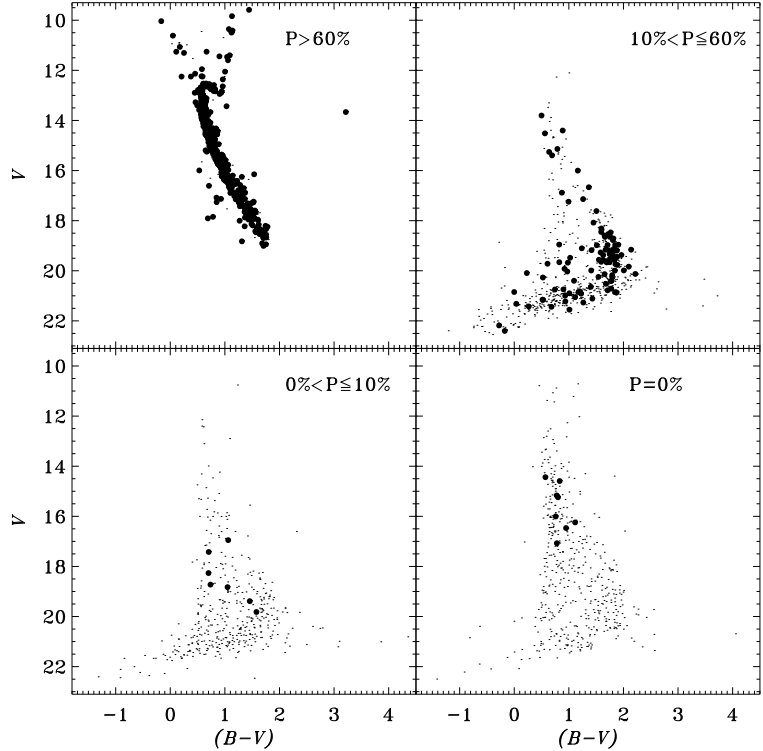}
\caption{
CMD-diagrams. 
Panels correspond to the mentioned membership probabilities $P$. 
Stars inside $\mu$-box are shown with bold circles.
}
\label{figu3}
\end{figure}

\begin{figure}[]
\centering
\includegraphics[width=8cm]{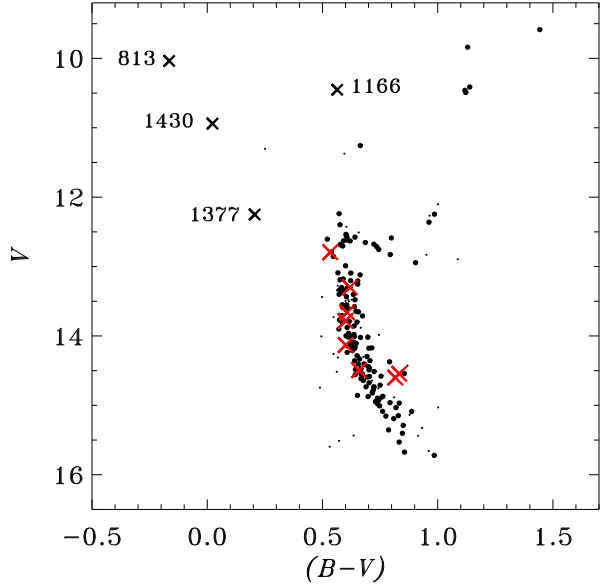}
\caption{Color-magnitude diagram for stars with measured $V_r$. 
The dots show all the stars, while the filled circles are stars inside AD-area. 
Red crosses represent the stars with $P>60$\% but located 
outside AD-area and out of $\mu$-box. 
}
\label{figu4}
\end{figure}

\begin{table}
\footnotesize
\caption{Data for the stars located far from the point of concentration in the AD-diagram.}
\label{tab6}
\vspace{0.5cm}
\centering
\tabcolsep=1.5mm
\begin{tabular}{c c c c c c c c c c}
\hline\hline
\noalign{\smallskip}
N&$P$&$\mu_U$&$\mu_T$&$\sigma_{\mu_U}$&$\sigma_{\mu_T}$& Inside &$\lambda$&$V_r$&$\sigma_{V_r}$\\
\noalign{\smallskip}
$[\#]$& \%& (mas/yr)&(mas/yr)&(mas/yr)&(mas/yr)&$\mu$-box&(deg)&(km/s)&(km/s)\\
\hline
\noalign{\smallskip}
813&98&5.42&-0.54&3.48&1.00&yes&163.07&-28.279&0.260\\
1166&84&-1.37&-6.68&13.90&3.45&no&81.30&4.026&0.024\\
1377&98&-5.54&-0.14&7.66&1.89&yes&173.13&-29.728&0.793\\
1430&96&14.78&1.16&8.80&2.49&yes&165.67&-51.449&0.762\\
\hline
\end{tabular}
\end{table}
The star numbers 813, 1377, 1430 have 
negative $V_r$ values ($-28.3$, $-29.7$, $-51.4$~km/s correspondingly). 
These stars are located above turn off point of the main sequence 
on the CMD (see Fig.~\ref{figu4}).
\citet{Liu2008} and \citet{Pribulla2008} have shown that these stars are blue stragglers.
On this basis, we do not reject their membership.

Star number 1166 has $V_r=4$~km/s, 
that is low in comparison to the mean heliocentric radial velocity 
$V_{rm}=34.7\pm8.6$~km/s for M~67 
estimated by us using 165 cluster members with $P>60\%$. 
We found no additional information for this star.
Taking into account small $V_r$-value, its position on the CMD (see Fig.~\ref{figu4}) 
and AD-(see Fig.~\ref{fig1}) diagrams, 
it more likely does not belong to the cluster. 
If this star exhibits variable characteristics, its membership status can be revised.

We have a sample of 169 stars with known $V_r$ values and membership probabilities $P>60\%$.
Four stars that are far from the maxima of star density in the AD-diagram were not included in 
this list. Hence, in total, we used 165 stars to calculate the cluster apex coordinates. 
We have calculated the apex position by averaging 
$V_x$, $V_y$ and $V_z$ components of the space velocities of 
the stars as was done in \citet{Chupina2001} too. The cluster apex coordinates are found as: 
$A_0=132.97^\circ\pm0.81^\circ$, $D_0=11.85^\circ\pm0.90^\circ$. 
The error in the estimation of apex's 
coordinates depends on the errors of parameters used in the calculation. 
We have used the constant distance value for all the stars 
because we do not have individual parallaxes.

We have calculated the mean angular distance between 
the cluster's apex and the individual star apexes.
This value is equal to $24^\circ$. To define this value, we used the stars, 
for which we can calculate the individual apexes: they have proper motion, 
radial velocity and an average parallax. Number of these stars is equal to 211. 
The stars, that deviate up to $24^\circ$ from the cluster's apex in 
the AD-diagram, have common direction of the space velocity 
within the errors and they are the most 
probable cluster members. Hereafter, this area is named as ``AD-area''.

We have 169 stars with measured $V_r$ and $P>60$\%. 
From these 169 stars, four stars have been rejected, which jump aside on the AD diagram in Fig.~\ref{fig1}. 
Hence, we are left with 165 stars with  measured $V_r$ and $P>60$\%,
Out of these, 162 stars are inside $24^\circ$, i.e. ``AD-area'' of the  AD-diagram
with various membership probability values.

\section{Membership list specification}

Using $A_0,\, D_0$, we have calculated $\mu_U,\, \mu_T$. 
The distance reduction of $\mu_U$ and $\mu_T$ 
is not possible due to lack of individual distances of the stars. 
We have corrected $\mu_U$ only for 
the $\lambda$ angle (between star and apex directions). Because of the compactness of M~67,
apex is almost coinciding with the cluster center (M~67 is moving along the radial line), 
so $\lambda_0$ is small and $\sin(\lambda_0)$ is almost zero. For ease of calculation, we 
used $\lambda_0=0.5^\circ$.
As this normalization factor, its value is not so important.


\begin{table}
\caption{Number of stars in different areas in ``$\mu_U-\mu_T$'' and AD-diagrams. 
$N$ is the number of stars while $P$ is the membership probability (in \%).}
\vspace{0.5cm}
\label{tab5}
\centering
\begin{tabular}{c c c c c c c}
\hline\hline
\noalign{\smallskip}
$\mu$-box&AD-area&$N$&$P=0$&$0<P\le10$&$10<P\le60$&$P>60$\\
\hline
\noalign{\smallskip}
\multicolumn{2}{c}{$\mu_U - \mu_T$ diagram}&2410&554&489&772&595\\
 in&&664&8&8&122&526\\
out&&1746&546&481&650&69\\
\hline
\multicolumn{2}{c}{AD-diagram}&211&26&9&7&169\\
 &in&150&0&3&0&147\\
&out&61&26&6&7&22\\
\hline
$\mu$-box&AD-area&$N$&$P=0$&$0<P\le10$&$10<P\le60$&$P>60$\\
\hline
in&in&142&0&0&0&142\\
in&out&13&0&0&1&12\\
out&in&8&0&3&0&5\\
out&out&48&26&6&6&10\\
\hline
\end{tabular}
\end{table}

The value of $\mu_T=0$ means that the proper motion vector is directed exactly toward the apex. The 
deviation value from this direction is important criterion for membership probability definition.
This criterion helps us for the selection of members. 
The vector modulus ($\mu_U$) can be used as an additional criterion. 
Note that we refine the membership list from Yadav08 and do not make a new one. To determine 
the range of $\mu_U$ and $\mu_T$ values that distinguish the cluster members, 
we have 169 stars with $P>60\%$ and we excluded 
four stars with numbers 813, 1430, 1377 and 1166. 
Thus, in total, we have 165 stars. 
We have calculated: $\langle\mu_T\rangle=-0.03$~mas/yr, $\sigma_{\mu_T}=1.78$~mas/yr and
$\langle\mu_U\rangle=-0.075$~mas/yr, $\sigma_{\mu_U}=6.75$~mas/yr. 
Taking into account the $2\sigma$, 
we obtain the box area: $-13.57 < \mu_U < 13.42$ and $-3.59 < \mu_T < 3.53$. 
Hereafter, this area is named as ``$\mu$-box''. 
The stars, that are inside $\mu$-box in ``$\mu_U-\mu_T$''-diagram, are the most 
probable cluster members.

Figure~\ref{fig2} shows the ``$\mu_U-\mu_T$''-diagrams for stars 
with different values of membership probabilities. 
The $\mu$-box is shown with a rectangular box inside each panel. 
CMDs for stars with different range of membership probability values is shown on Fig.~\ref{figu3}, 
where the stars inside $\mu$-box are signed with solid circle.
The number of stars with various probabilities are presented in Table~\ref{tab5}.

In Fig.~\ref{fig2}, we see that there are stars outside $\mu$-box, 
but they have a large value of membership probability 
according to Yadav08 catalogue. 
Also on the top left panel on Fig.~\ref{fig2}, we notice a number of stars outside $\mu$-box. 
Since they are selected as the most probable ($P>60$\%) by Yadav08 
by means of CMD and proper motions diagram. 
The question of these stars are being outside $\mu$-box remains open.
The radial velocities of these stars can serve to refine their membership status. Some of them can be short period variables or blue stragglers as well.

Also, a few stars with zero probability are inside $\mu$-box. 
Stars with $P=0$\% (right bottom panel) should not be inside $\mu$-box 
because they are strongly recommended to be nonmembers. 
On Fig.~\ref{figu3}, these stars are lying on the Main Sequence of the cluster. 
This can be the reason behind their belonging to the cluster. 
They cannot be checked using AD-diagram  
because these stars do not have $V_r$-values.
The status of their membership can be checked on the availability of better data.

In the bottom panels on Fig.~\ref{fig2}, an overdensity 
of stars near $\mu_U=+15$~mas/yr can be noticed.
Most likely, it is the consequence of the two factors. 
First possibility is that it can be due to features of cluster spatial movement to an apex direction
(remember that the axis $\mu_U$ is directed from the cluster center to an apex point, 
it witnesses an advancing of field stars proper motions in this direction). 
The second factor behind the overdensity can be the real half-width 
of the $\mu$-box being less than $2\sigma$.
This case can be seen in top left panel on Fig.~\ref{fig2} ($P>60$\%),
where size of the central points concentration is less than $\mu$-box limits.

It is useful to compare the list of the cluster members, 
which have been selected by $\mu$-box and AD-area methods. 
In total, 211 stars are studied by these two methods. 
Out of 211 stars, 190 are identically marked 
by both techniques in which 142 are members and 48  are non-members. 
The other 21 stars are found as contamination 
in which 8 are members by AD-area method and 
non-members by $\mu$-box method while 13 stars are found in opposite sense. 
The membership determination with the proper motion method 
is less reliable than AD-method, 
since it makes use of only two components of the space vectors 
and does not consider $V_r$. 
Anyway, we have $\sim$90\% of agreement between the two methods.

Table~\ref{tab5} shows that there are stars, 
that are selected by ``$\mu_U-\mu_T$''-criterion, but they are outside 
the AD-area, and vice versa. However, most of the stars satisfy both criteria. 
Since $V_r$ values are present for few stars, therefore, 
for most of the stars we define the membership by proper 
motion and make some comments based on the AD-diagram. 

\begin{figure*}[ht]
\centering
\includegraphics[width=14cm]{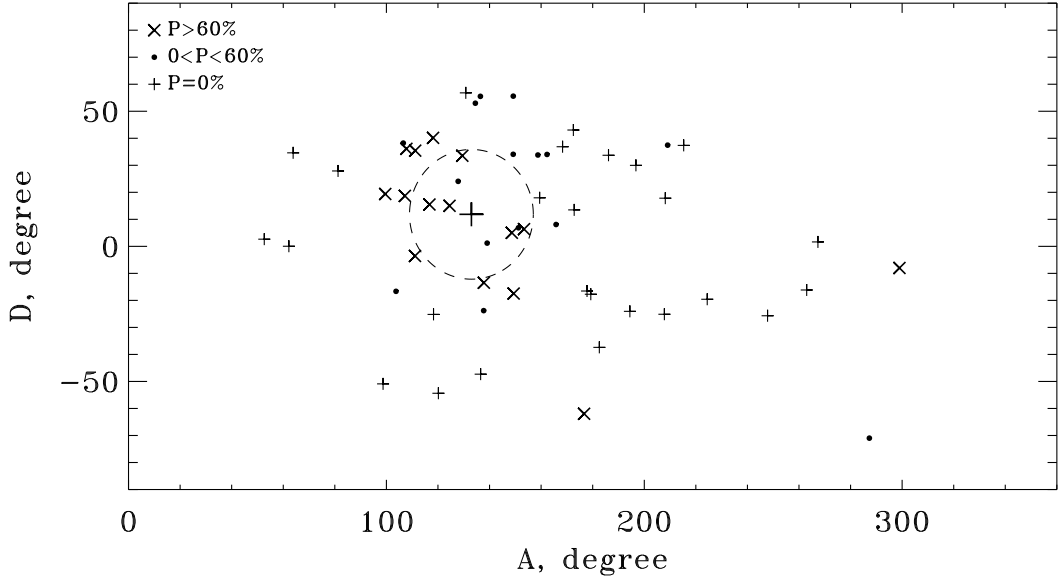}
\caption{The AD-diagram for stars outside $\mu$-box for different membership probabilities
as shown in the top left corner.}
\label{figu5}
\end{figure*}

Now, we check the stars in AD-diagram that are outside $\mu$-box shown in Fig.~\ref{figu5}. 
We see that the stars with $P=0\%$ and outside $\mu$-box,
confirm their non membership by AD-area too. 
22 stars with $P>0\%$ are outside both $\mu$-box and AD-area. 
The status of their membership needs revision.
One of them is 1166, about which we have already discussed.
There are stars with $P>0\%$, 
that are outside $\mu$-box, but inside AD-area. 
Since we consider AD-method more solid, 
we do not revise their status of membership.

\subsection {The general comments on membership}

The rectangular area has been defined on ``$\mu_U-\mu_T$''-diagram shown in Fig.~\ref{fig2} by 
means of proper motion statistics.
The most probable members of the cluster are located inside this area. 
Most of the stars marked in the 
Yadav08 catalogue as probable members, are within this area, and thus confirmed their membership. 
However, there are disagreements too. 
Present results should be considered as the supplement to the Yadav08 
catalogue. There will be numbers of $P$ values from Yadav08 and $\mu$-flag from our study. The 
$\mu$-flag equals to 1 or 0 if star is inside or outside the $\mu$-box respectively.

We would like to comment on the membership of some stars due to their positions in the AD-diagram.
These are the stars 1166, 813, 1377 and 1430, which deviate from the cluster apex owing to 
the radial velocity values. These radial velocity values needs to be checked. Figure~\ref{figu4} 
shows that these stars are not falling on cluster sequence and may be non-members. 
Other eight stars with numbers 401, 548, 930, 1088, 1089, 1480, 1716, 1722 
keep well within the cluster main sequence shown on Fig.~\ref{figu4} and have $P>60$\%. 
However, they hold positions out of the $\mu$-box on Fig.~\ref{figu5} and out of AD area on Fig.~\ref{fig2}.
Thus, the listed eight stars do not belong to the cluster more likely.

\begin{figure}
\centering
\includegraphics[width=7cm,height=7cm]{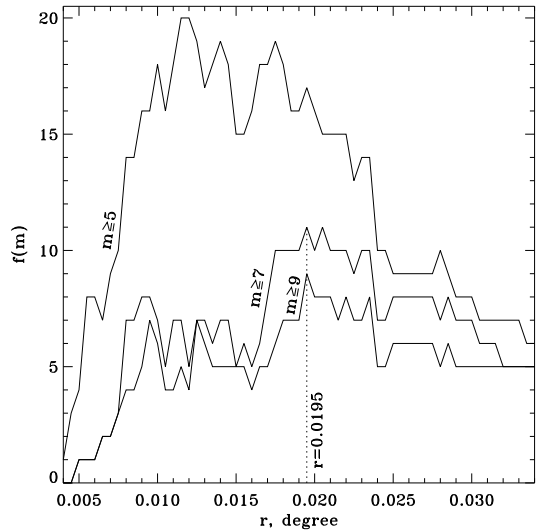}
\caption{The number of clumps, $f(m)$, depending on distance between neighboring stars $r$. 
The maximum of the curve indicates ``characteristic scale'' $r_0$, 
which leads to a clustering of the groups
including $m$ and more stars in the structure. 
Dotted vertical line shows $r_0=0.0195^\circ$ and 
has been used for group allocation.}
\label{figu6}
\end{figure}

\section{Substructures}

Nearest neighbour distances (hereafter, NND) method \citep{Einasto1984, Battinelli1991}   
has been used for discovering spatially bounded structures. 
This method is useful for the isolation 
of structures of different sizes and configurations. 
In essence: according to the NND method, 
star belongs to the group if its distance to the nearest star from the group 
does not exceed a certain value $r_0$. 
There are different clustering patterns for different $r_0$. 
For instance, if $r_0$ is greater than any interstellar distance, 
then all stars are clumped. 
On the other hand, if $r_0$ has a minimal possible value, 
only one group of two stars will be allocated. 
Between two extreme values of $r_0$ there is one which results 
to the maximum possible number of groups 
with different number of stars in them. 
This is a characteristic scale for studying the inner structures 
in the stellar system.

Use of NND method for M~67 has specified the heterogeneity of both coronae and nucleus
using the distribution of the number of groups by characteristic distance 
between two neighboring stars in groups (Fig.~\ref{figu6}). 
The maxima of the curve defines different scales of 
substructures. For example, for the center and the periphery.

First of all, we have determined the characteristic scale $r_0$, 
i.e. the maximum possible distance between 
two neighboring stars of the clump \citep{Chupina1998}. 
We counted the number of clumps, 
$f(m)$ based on different distances (see Fig.~\ref{figu6}), 
where a clump is a group of $m$ or more stars. 
We used stars with $P>60\%$ and got $r_0=0.0195^\circ$ ($1.17'$).


\begin{table}
\caption{Parameters of the clumps in the present study.}
\label{tab7}
\vspace{0.5cm}
\centering
\tabcolsep=1.5mm
\begin{tabular}{c  c c c}
\hline\hline
\noalign{\smallskip}
Analysed portion &$r_0$&Diameter&Number of stars\\
of  M~67 & & of clump& in groups  \\
\hline
\noalign{\smallskip}
Nucleus& 1.17$^\prime$& 0.1 deg&5 -- 17 \\
\hline
\end{tabular}
\end{table}

\begin{figure}
\centering
\includegraphics[width=9cm,height=9cm]{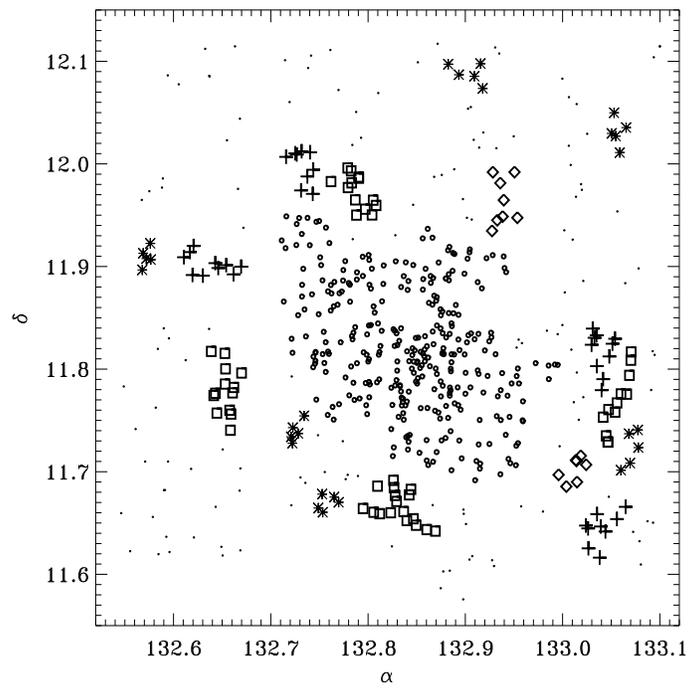}
\caption{The stars distribution in the equatorial coordinate system. 
Different symbols designate the 
groups of stars obtained by NND-method.}
\label{figu7}
\end{figure}

With this $r_0$-value, 
we have detected the clumps among stars with $P>60\%$ (see Fig.~\ref{figu7}).
Parameters of the clumps are listed in Table~\ref{tab7}. 
It is seen that the central part of the 
cluster is homogeneous (filled circles in Fig.~\ref{figu7}).
Its radius is $\approx0.1^\circ$. 
This is in agreement with the estimate of the core radius from \citet{Kharchenko2005}. 
A more distant region is fragmented (points shown with different symbols in 
Fig.~\ref{figu7}). There are several groups which include five or more stars.

If we use other sample of stars (for example, with $P>0\%$), 
we obtain another $r_0$-value, 
but the clumps are detected at the same places with different number of stars. 
With a decrease in the $r_0$-value, 
the number of clumps and the number of stars in clumps are increased, and vice versa.

\citet{Chupina1998} used smaller sample that led to bigger $r_0$ ($4.0'$) 
and a little less number of groups were allocated (5 against 16 in this work). 
The group sizes and the number of stars in them are very similar. 
An exact comparison is impossible due to the different areas covered. 
This study is restricted to the central area 
but deeper data while \citet{Chupina1998} studied the corona up to $V$ around $16$ mag. 

The spaces between CCDs in WFI@2.2m are the star free regions 
seen in Fig.~\ref{figu7}. The value used for $r_0=0.0195^\circ$ 
is much less than the width of these bands. 
This means that the stellar population, located in blanks, may somewhat change the result. 
So the results can be improved with wider area of observation 
with no gaps in the detector.

%

\begin{figure}
\centering
\includegraphics[width=7cm,height=7cm]{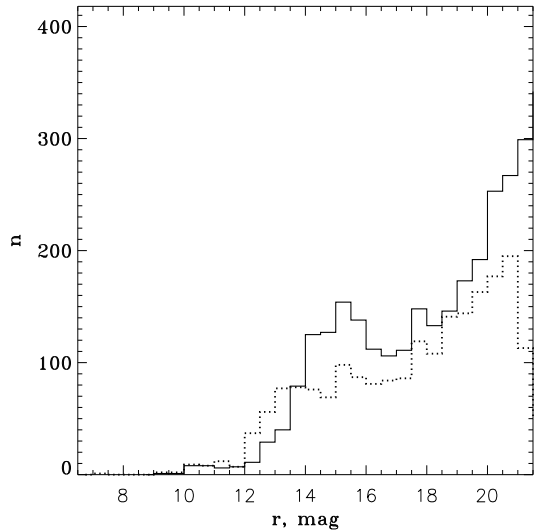}
\caption{Distribution in r-magnitude by SDSS data (solid line, 2375 stars)
and Yadav08 data (dotted line, 1917 stars).
}
\label{fig8}
\end{figure}

For the test of completeness Of stars in Yadav08 catalogue,
the homogeneous photometry was taken 
from the Sloan Digital Sky Survey\footnote{http://cas.sdss.org/dr7/en/tools/search/form/form.asp} (SDSS). 
The comparison was done in the same observational area and a histogram is shown in Fig.~\ref{fig8}. 
The comparison was made in $r$-magnitude of the SDSS photometric system. 
For the comparison purpose, we converted the $V$ magnitude 
of Yadav08 catalogue to the SDSS $r$-magnitudes
with the formula 
$r=V-0.46(B-V)+0.11$\footnote{http://www.sdss.org/dr5/algorithms/sdssUBVRITransform.html}.

The histograms of the converted $r$-magnitudes and SDSS $r$-magnitudes are given in Figure~\ref{fig8}.
For the histograms, we have 1917 stars from Yadav08 catalogue and 2375 stars from SDSS up to $r_{lim}=21$ mag.
The difference between the two samples is of 458 stars.
Therefore, 19\% loss of stars in Yadav08 catalogue may be due to the blanks on mosaic CCDs.
However, there is a surplus of stars in the Yadav08 catalogue in the $r$-magnitude bin of 12--13 mag.
This can be due to the photometry in the brighter end of Yadav08 being better than the SDSS.

%


\section{Conclusions}

From the present analysis for M~67 open cluster, the following main conclusions can be drawn: 

\begin{enumerate}

\item The apex coordinates for M~67 have been calculated as:
$A_0=132.97^\circ\pm0.81^\circ,\, D_0=11.85^\circ\pm0.90^\circ$.

\item The membership of the stars has been revisited. In future, it is possible to decide membership by 
distance from $\mu$-box and by position of the star in ``$\mu_U-\mu_T$''-diagram. If $\mu_T$ is 
close to zero, the membership probability is higher (within errors and peculiar velocities dispersion).

\item AD-diagram has been constructed for stars with known radial velocities. The membership status for the 
star can be estimated with the help of AD-diagram. The farther away a star located from the apex 
position, there will be the less probability for it to be a cluster member.

\item We found substructures in the corona of M~67.
We found heterogeneity in the core periphery, like our  previous work \citep{Chupina1998}.

\end{enumerate}

\section{Acknowledgments}

Authors are highly thankful to an anonymous referee for a careful reading of the draft
and providing with useful scientific and constructive comments.
This work is done under a joint research collaboration project, 
titled "Photometric and kinematical studies 
of the galactic disk population" 
between Russia and India. 
The project code is INT/ILTP/B-3.21 (2010-2013). \\

\bibliographystyle{model2-names}



\end{document}